\documentclass[11pt]{llncs}
\usepackage{epsfig}
\usepackage{graphics}
\usepackage{amsmath,setspace}

\usepackage[dvipsnames,usenames]{color}
\usepackage[colorlinks=true,urlcolor=Blue,citecolor=Green,linkcolor=BrickRed]{hyperref}
\begin{document}

\title{Binary Jumbled Pattern Matching\\ via All-Pairs Shortest Paths}

\author{Danny Hermelin \inst{1}\thanks{Supported in part by the Marie Curie Career Integration Grant (CIG) 631163.}  \and Gad M. Landau\inst{2}\thanks{Supported in part by the National Science Foundation (NSF) grant 0904246, the Israel Science Foundation (ISF) grant 347/09,
and the United States-Israel Binational Science Foundation (BSF) grant 2008217.}  \and \\ Yuri Rabinovich\inst{2}
\and Oren Weimann\inst{2}\thanks{Supported in part by the Israel Science Foundation (ISF) grant 794/13.}}
\institute{
Ben-Gurion University, {hermelin@bgu.ac.il} \and
University of Haifa, \{landau,yuri,oren\}@cs.haifa.ac.il}

\maketitle

\begin{abstract}
In binary jumbled pattern matching we wish to preprocess a binary string $S$ in order to answer queries $(i,j)$ which ask for a substring of $S$ that is of size $i$ and  has exactly $j$ 1-bits. The problem naturally generalizes to node-labeled trees and graphs by replacing ``substring'' with ``connected subgraph''.

\hspace{15pt} In this paper, we give  ${n^2}/{2^{\Omega(\log n/\log \log n)^{1/2}}}$ time solutions for both strings and trees. This odd-looking time complexity improves the state of the art $O(n^2/\log^2 n)$ solutions by more than any poly-logarithmic factor. It originates from the recent seminal algorithm of Williams for min-plus matrix multiplication.
We obtain the result by giving a black-box reduction from trees to strings. This is then combined with a  reduction from strings to min-plus matrix multiplications.
\end{abstract}

\section{Introduction}
\label{sec:introduction}

A string $P$ is said to have a {\em jumbled} occurrence in string $T$ if $P$ can be rearranged so that it appears in $T$. In other words, if $T$ contains a substring of length $|P|$ where each letter of the alphabet occurs the same number of times as in $P$.
In indexing for {\em Jumbled pattern matching} we wish to preprocess a given text $T$ so that given a query $P$ we can determine quickly whether $T$ has a jumbled occurrence of $P$.


\paragraph{\bf Binary jumbled pattern matching on strings.}
Apart from a recent paper on constant alphabets~\cite{ESA13}, all the
 results on the problem are restricted to binary alphabets (where a query pattern $(i,j)$ asks for a substring of $T$ that is of length $i$ and has $j$ 1s). The important property of a binary alphabet is that $(i,j)$ appears in $T$ iff $j$ is between the minimum and maximum number of 1s over all substrings of length $i$.
As observed in~\cite{CFL09}, this means that we can store only the  minimum and maximum values of every $i$ and can then answer a query in $O(1)$ time. While this requires only $O(n)$ space, computing it naively takes $O(n^2)$ time. Beating $O(n^2)$ has become a recent challenge of the pattern matching community.

The first improvement was to $O(n^2 / \log n)$. It was independently obtained by 
Burcsi et al.~\cite{BCFL12a}, and by Moosa and Rahman~\cite{MR10}, who reduced the problem to min-plus products of vectors. Moosa and Rahman~\cite{MR12} then further improved it to $O(n^2 / \log^2 n)$ in the RAM model by cleverly using the four-Russians technique instead of min-plus products. This remained the state of the art and $o(n^2 / \log^2 n)$ time was only known when the string compresses well under run-length encoding~\cite{BFKL12,GG12} or when we are willing to settle for approximate indexes~\cite{CLWY12}.

 \paragraph{\bf Binary jumbled pattern matching on Trees.}
On a tree $T$ whose nodes are labeled 0 or 1, a query $(i,j)$ asks for a connected subgraph of $T$ that is of size $i$ and has exactly $j$ nodes labeled by 1. Like in strings, if $(i,j_1)$ and $(i,j_2)$ both appear in $T$, then for every $j_1\le j \le j_2$,  $(i,j)$ appears in $T$. This means that, again, we only need to store for every $i$ the minimum $j_1$ and maximum $j_2$ values such that  $(i,j_1)$ and $(i,j_2)$ appear  in $T$.
In~\cite{OurESA13} we showed that finding these values can be done in $O(n^2 / \log^2 n)$ time, just like in strings.  In fact, the solution for trees was obtained by reducing it to multiple applications of the solution for strings~\cite{MR12} based on the four-Russians technique.

\paragraph{\bf Our results.} Given a string (resp. tree) $T$, we refer to jumbled pattern matching as the problem of computing for every $i$ the maximum and minimum number of 1s in a substring (resp. connected subgraph) of $T$ of size $i$.
 We obtain the following:

\begin{theorem}\label{theorem1}
Any $O(n^3/\ell(n))$ algorithm for computing the min-plus product of $n\times n$ matrices implies an
 $O(n^2/\ell(\sqrt{n}))$ algorithm for jumbled pattern matching on strings and an $O(nr+ n^2/\ell(\sqrt{r}))$-time algorithm for  jumbled pattern matching on trees for any choice of $r$.
\end{theorem}

\noindent Our work was motivated by the recent breakthrough algorithm of Williams~\cite{Ryan} for computing the min-plus product of two $n\times n$ matrices in $n^3/2^{\Omega(\log n/\log \log n)^{1/2}}$ time\footnote{Using the Williams algorithm on the word RAM takes $n^3/2^{\Omega(\log n/\log \log n)^{1/2}}$ time since we know that all elements of our matrices are bounded by $\sqrt{n}$. The algorithm is randomized and can be made deterministic  in $O(n^3/2^{\log^{\delta} n})$ time for some $\delta > 0$.}. This means that currently both $\ell(n)$ and  $\ell(\sqrt{n})$ are $2^{\Omega(\log n/\log \log n)^{1/2}}$ (the difference between $\ell(n)$ and  $\ell(\sqrt{n})$ is only in the constant behind the $\Omega$). Choosing $r= \sqrt{n}$, we get:

\begin{corollary}\label{cor}
Jumbled pattern matching on both strings and trees can be solved in $n^2/2^{\Omega(\log n/\log \log n)^{1/2}}$ time.
 \end{corollary}

\noindent Finally, we note that the above bound also applies to the more general problem of computing the maximum sub-sums  of a string or a tree. Namely, given a string (resp. tree) whose characters (resp. nodes) have arbitrary weights we can compute (in the time bound of Corollary~\ref{cor}) for every $i=1,\ldots,n$ the maximum sum of weights of all substrings (resp. connected subgraphs) of size $i$.

\section{Binary Jumbled Pattern Matching on Strings}

We begin by proving the first part of Theorem~\ref{theorem1} regarding jumbled pattern matching on strings. As discussed in the previous section, this boils down to the following problem: Given a binary text $T$ of length $n$, compute the minimum and maximum number of 1s in a substring of length $s$ in $T$, for all $s=1,\ldots,n$. Below, we focus on computing the minimum number of 1s in each substring length, as computing the maximum number of 1s can be done in an analogous manner. We show how to do this in total $O(n^2/\ell(\sqrt{n}))$ time, where $\ell(n)$ is the assumed speedup factor for the naive cubic-time min-plus multiplication algorithm of matrices $A$ and $B$, defined as:  \[(A \star B)[i,j] = \min_{k} (A[i,k] + B[k,j]).\]
That is, matrix multiplication where $\min$ plays the role of addition, and $+$ plays the role of multiplication.
The  complexity of  such multiplication is equivalent to that of All-Pairs Shortest Paths.

We start by first partitioning the string $T$ into consecutive substrings (blocks) $T_0,\ldots, T_{\sqrt{n}-1}$ each of length $\sqrt{n}$. We then compute for every $T_i$ the minimum number of 1s in a substring of length $s$ that is completely inside~$T_i$. This can be done naively for all $s\in\{1,\ldots, \sqrt{n}\}$ in $O(n)$ time, and over all $T_i$'s in $O(n^{1.5})$ time.

We next want to compute the minimum number of 1s in substrings that span more than one block. For every $\ell \in \{1,\ldots, 2\sqrt{n}\}$, let $C_\ell$ be the $\sqrt{n}\times \sqrt{n}$  matrix where $C_\ell[i,j]$ is the minimum number of 1s in substrings that include: (1) a suffix $q$ of $T_i$ (2) the complete blocks $T_{i+1},\ldots,T_{j-1}$ (3) a prefix $p$ of $T_j$, and (4) $\ell=|p|+|q|$. It is not hard to see that once we have all $C_1,\ldots,C_{2\sqrt{n}}$, along with all information we computed within the blocks, solving our problem is trivial in $O(n^{1.5})$ time.

We distinguish between two cases: The case where $\ell \leq \sqrt{n}$ (in which $p$ and $q$ are allowed to be empty), and the case where $\ell > \sqrt{n}$ (in which both $p$ and $q$ must be non-empty).

Assume that $\ell \le \sqrt{n}$. Let $A$ be the  $\sqrt{n}\times (\ell+1)$ matrix such that $A[i,k]$ is the number of 1s in the last $k$ bits of $T_i$. Similarly, we define the $(\ell+1) \times \sqrt{n}$ matrix $B$ such that $B[k,j]$ is the number of 1s in the first $\ell-k$ bits of $T_j$. Their min-plus product $C$ is defined as $C[i,j] = \min_{k} (A[i,k] + B[k,j])$. We set $C_\ell[i,j] = C[i,j]$ + $x_{i,j}$ where $x_{i,j}$ is the number of 1s in the substring $T_{i+1} \cdots T_{j-1}$. Note that computing $x_{i,j}$ is done once and is then used for every $\ell$.

To compute $C_\ell$ for $\ell > \sqrt{n}$, we use the same procedure and only slightly change $A$ and $B$. Now $A$ is an  $\sqrt{n}\times (2\sqrt{n}-\ell+1)$ matrix, and $A[i,k]$ is the number of 1s in the last $k+\ell-\sqrt{n}$ bits of $T_i$. The matrix $B$ is an~$(2\sqrt{n}-\ell+1)\times \sqrt{n}$  matrix such that $B[k,j]$ equals the number of 1s in the first $\sqrt{n}-k$ bits of $T_j$. The matrix $C$ is again defined as the min-plus product $A \star B$, and $C_\ell[i,j]$ is computed as in the previous case.

Note that computing $A$ and $B$ for each $\ell$ can be trivially done in $O(n)$ time. Furthermore, it is not difficult to see that  the value $C_\ell[i,j]$ computed for each $i$ and $j$ is indeed the minimum number of 1s in substrings of $T$ as required above. The matrix $C_\ell$ can be computed easily in $O(n)$ time once $C$ has been computed via the min-plus computation. Since $\ell=O(\sqrt{n})$, using the algorithm of Williams, we can compute this product in $O(n^{3/2}/\ell(\sqrt{n}))$ time. Thus, in total we compute $C_1,\ldots,C_{2\sqrt{n}}$ in $O(n^2/\ell(\sqrt{n}))$ time. This proves the first part of Theorem~\ref{theorem1}.

\subsection{Relation to Previous Work}
The above proof was first suggested by us in 2008~\cite{Private2008}. A similar construction was independently obtained by Bremner et al.~\cite{BCDEHILT06} (Arxiv 2012, Section 4.4) who showed that MPV$(n)= n^{1.5} + \sqrt{n}\cdot $MPM$(\sqrt{n})$.  Here, MPM$( \sqrt{n})$ denotes the time it takes to compute the min-plus product of two $ \sqrt{n}\times  \sqrt{n}$ {\em matrices} and  MPV$(n)$ denotes the time it takes to compute the min-plus product of two $n$-length {\em vectors}  $x,y$
defined as: \vspace{-0.07in} \[(x \odot y)[i] = \min_{k=1}^i (x[k]+y[i-k]).\]

Moosa and Rahman~\cite{MR10} showed that jumbled pattern matching on a string of length $n$  can be done in time $T(n) = 2T(n/2) +$MPV$(n)$. By Bremner et al. this means that $T(n) = 2T(n/2) +n^{1.5} + \sqrt{n}\cdot $MPM$(\sqrt{n})$. By
Williams~\cite{Ryan} we have MPM$(\sqrt{n})= O(n^{3/2}/\ell(\sqrt{n}))$ and so $T(n)=O(n^2/\ell(\sqrt{n}))$.

\section{Binary Jumbled Pattern Matching on Trees}

We now prove the second part of Theorem~\ref{theorem1}. Given a tree $T$ with~$n$ nodes, each labeled with either 0 or 1, we wish to compute, for every $i=1.\ldots, n$ the minimum number of nodes labeled 1 in a connected subgraph of $T$ that is of size $i$ (the maximum is found similarly). In~\cite{OurESA13}, we presented a tree-to-strings reduction for this problem that was based on the four-Russians speedup of~\cite{MR12}. Here, we generalize this reduction to a black-box reduction, which is applied regardless of the particular speedup technique used in the string case. We outline this generalization below.

The first observation in~\cite{OurESA13} was that we can assume w.l.o.g that $T$ is a binary tree. The second was an $O(n^2)$ simple algorithm: In a bottom-up manner, for each node $v$ of $T$, compute an array $A_v$ of size $|T_v|+1$ ($T_v$ includes $v$ and all its descendants in $T$). The entry $A_v[i]$ will store the minimum number of 1-nodes in a connected subgraph of size $i$ that includes $v$ and another $i-1$ nodes in $T_v$. If $v$ has a single child $u$, then we set $A_v[i]=lab(v)+A_u[i-1]$, where $lab(v)$ is the label of $v$. If $v$ has two children $u$ and $w$,  we set $A_v[i]= lab(v)+\min_{0 \leq j \leq i-1} \{A_u[j]+A_w[i-j-1]\}$. The time required to compute all arrays is asymptotically bounded by $\sum_v \alpha(v)\beta(v)= O(n^2)$ where $\alpha(v)$ (resp. $\beta(v)$) is the size of $v$'s left (resp. right) child's subtree.

The total space used can be made $O(n)$ by only keeping $A_v$'s which are necessary for future computations. It can be made $O(n)$ {\em bits} by representing $A_v$ as a binary string $B_v$ where $B_v[0]=0$, and $B_v[i]= A_v[i]-A_v[i-1]$ for all $i=1,\ldots,n-1$. Since $A_v[i]= \sum_{j=0}^i B_v[j]$, each entry of $A_v$ can be retrieved from $B_v$ in $O(1)$ time using {\em rank} queries~\cite{Jacobson}.

\paragraph{\bf The black-box reduction.}

Now that we have an $O(n^2/\ell(\sqrt{n}))$-time algorithm for  strings we would like to also obtain an $O(n^2/\ell(\sqrt{n}))$-time algorithm for trees. Using the above algorithm, this can be achieved if computing $B_v$ can be done in $O(\alpha(v)\beta(v)/\ell(\sqrt{n}))$ time, since in total we would then get $O(1/\ell(\sqrt{n}) \cdot \sum_v \alpha(v)\beta(v)) = O(n^2/\ell(\sqrt{n}))$ time.

For a node $v$ with children $u$ and $w$, we can compute $B_v$ using jumbled pattern matching on the binary string $S = X \cdot lab(v) \cdot Y$, where $X$ is obtained from $B_u$ by reversing it and removing its last bit, and $Y$ is obtained from $B_w$ by removing its first bit. The catch is that we are only interested in substrings that include the position of $lab(v)$ in $S$. If $x=|X|$ and $y=|Y|$, then this can naively be done in $O(xy)$ time. Alternatively, it can also be done in $O(|S|^2/\ell(\sqrt{|S|}))= O((x+y)^2/\ell(\sqrt{x+y}))$ time using the algorithm of the previous section\footnote{Note that the algorithm from the previous section can easily be adapted (in the same time complexity) to only consider substrings that include the position of $lab(v)$.}. However, we desire $O(xy/\ell(\sqrt{n}))$.

To achieve this, assume w.l.o.g that $x \leq y$. We partition $Y$ into consecutive substrings $Y_1,\ldots,Y_{y/x}$, each of length $x$ (except perhaps the last one). We compute $B_v$ by solving jumbled pattern matching on all the strings  $X \cdot lab(v) \cdot Y_i$. Using the previous section this takes total $(y/x)\cdot (x^2/\ell(\sqrt{x}))= O(xy/\ell(\sqrt{x}))$. This would be fine if $\ell(\sqrt{x})$ is roughly equal to $(\ell(n))$. Note that for large enough $x$, say $x\!\ge\!\! \sqrt{n}$, with the current Williams bound  indeed both $\ell(\sqrt{x})$ and $\ell(n)$ are $2^{\Omega(\log n/\log \log n)^{1/2}}$.  The challenge is therefore  to deal with small $x$ (say $x\!< \!\sqrt{n}$).

The challenge of small $x$ was also an obstacle in the reduction of~\cite{OurESA13}.
We use the same solution of~\cite{OurESA13} (with only a small change in parameters). Namely a \emph{micro-macro decomposition}~\cite{MicroMacro}.
A micro-macro decomposition is a partition of $T$ into
$O(n/r)$ disjoint connected subgraphs of size at most $r$ called {\em micro trees}. Each micro tree $C$ has at most two nodes (called \emph{boundary} nodes) that are adjacent to nodes in other micro trees. The {\em macro tree}  is a tree of size $O(n/r)$. Each node of the macro tree corresponds to a micro tree $C$ and the edges of the macro tree  to edges between boundary nodes.

%

We first compute the maximum number of 1s in all patterns that are completely inside a micro tree. Using the above simple algorithm each micro tree is handled in $O(r^2)$ time, so overall $O((n/r)\cdot  r^2)=O(nr)$. Notice that in particular this computes $B_v$ for every boundary node $v$ with respect to its micro tree $C$. Denote this array by $B_v(C)$.

To deal with patterns that span multiple micro trees, it was shown in~\cite{OurESA13} that the simple algorithm can be applied bottom-up on the macro tree (instead of on $T$). For each node $C$ in the macro tree, and each boundary node $v$ of $C$, the array $B_v$ is computed by combining the array $B_v(C)$ with the arrays $B_u$ of every descendant boundary node $u$ adjacent to $v$. Recall that combining the arrays means solving jumbled pattern matching on a binary string $S = X\cdot lab(v) \cdot Y$ with $x=|X|$ and $y=|Y|$. As before, if $x\ge r$ this takes $O(xy/\ell(\sqrt{x}))=O(xy/\ell(\sqrt{r}))$ time, so over all such computations take $O(n^2/\ell(\sqrt{r}))$ time. If~$x < r$, we simply extend $x$ artificially until it is of length $r$. The computation will then take $O(ry/\ell(\sqrt{r}))=O(rn/\ell(\sqrt{r}))$ time, but there are only $O(n/r)$ boundary nodes, so overall this takes $O(n^2/\ell(\sqrt{r}))$ time. Accounting also for the $O(nr)$ time required for computing all necessary information inside the micro trees, we obtain the time complexity promised in Theorem~\ref{theorem1}.


\section{Conclusions}
We have showed that any $O(n^3/\ell(n))$ algorithm for computing the min-plus product of $n\times n$ matrices implies an $O(n^2/\ell(\sqrt{n}))$-time algorithm for  jumbled pattern matching on strings, and an $O(nr+ n^2/\ell(\sqrt{r}))$-time algorithm for jumbled pattern matching on trees for any choice of $r$. With the current Williams bound on $\ell(n)$, and by choosing $r=\sqrt{n}$, we get that jumbled pattern matching on either strings or trees can be done in $O(n^2/\ell(\sqrt{n}))$ time. This is because currently both $\ell(\sqrt{n})$ and $\ell(n)$ are $2^{\Omega(\log n/\log \log n)^{1/2}}$.

In the future, if say an $O(n^{3-\varepsilon})$ algorithm is found (with a constant $\varepsilon$) for All-Pairs Shortest Paths (i.e., for min-plus products), then we would get an $O(n^{2-\varepsilon/2})$ algorithm for jumbled pattern matching on strings but only an $O(n^{2-\varepsilon'})$ algorithm for trees where $\varepsilon' = \frac{\varepsilon/2}{1+\varepsilon/2}$.  This is obtained using the $O(nr+ n^2/\ell(\sqrt{r}))$ bound of Theorem~\ref{theorem1} with $r=n^\frac{1}{1+\varepsilon/2}$. However, notice that the $O(nr)$ factor originated from running the simple $O(r^2)$ algorithm on each one of the $O(n/r)$ micro trees. But we now have a better than $O(r^2)$ algorithm, namely an $O(rr'+ r^2/\ell(\sqrt{r'}))$ algorithm for any choice of $r'$. Doing this recursively improves the $O(nr)$ factor and makes $\varepsilon'$ closer to $\varepsilon/2$. We leave this as an exercise for the optimistic future in which APSP can be done in $O(n^{3-\varepsilon})$ time.

%
%

\bibliographystyle{plain}
\bibliography{jumbled}
\end{document}